\documentclass[prd,floatfix,preprintnumbers,letterpaper,twocolumn]{revtex4}
\usepackage{amsmath,amssymb,graphicx}
\usepackage{hyperref}

\newcommand{\lk}{\left}
\newcommand{\rk}{\right}

\begin{document}
\title{Testing the Etherington's distance duality relation at higher
redshifts: the combination of radio quasars and gravitational waves}
\author{Jing-Zhao Qi$^{1,2}$, Shuo Cao $^{2\ast}$, Chenfa Zheng$^{2}$, Yu Pan$^{3}$, Zejun Li$^{2}$, Jin Li$^{4\dag}$, and Tonghua Liu $^{2}$ }

\affiliation{1. Department of Physics, College of Sciences,
Northeastern University, Shenyang 110004, China;\\
2. Department of Astronomy, Beijing Normal University,
100875, Beijing, China; \emph{caoshuo@bnu.edu.cn}\\
3. College of Science, Chongqing University of Posts
and Telecommunications, Chongqing 400065, China;\\
4. Department of Physics, Chongqing University,
Chongqing 400030, China; \emph{cqujinli1983@cqu.edu.cn}}

\begin{abstract}
In this paper we analyse the implications of the latest cosmological
data sets to test the Etherington's distance duality relation (DDR),
which connects the luminosity distance $D_L$ and angular diameter
distance $D_A$ at the same redshift. For $D_L$ we consider the
simulated data of gravitational waves from the third-generation
gravitational wave detector (the Einstein Telescope, ET), which can
be considered as standard candles (or standard siren), while the
angular diameter distances $D_A$ are derived from the newly-compiled
sample of compact radio quasars observed by very-long-baseline
interferometry (VLBI), which represents a type of new cosmological
standard ruler. Alleviating the absorbtion and scattering effects of
dust in the Universe, this will create a valuable opportunity to
directly test DDR at much higher precision with the combination of
gravitational wave (GW) and electromagnetic (EM) signals. Our
results show that, with the combination of the current radio quasar
observations, the duality-distance relation can be verified at the
precision of $10^{-2}$. Moreover, the Einstein Telescope ET would
produce more robust constraints on the validity of such distance
duality relation (at the precision of $10^{-3}$), with a larger
sample of compact milliarcsecond radio quasars detected in future
VLBI surveys.
\end{abstract}

\maketitle

\section{Introduction} \label{introduction}

Due to the rapid technological advances in observational cosmology,
the accumulation of observational data obtained with increasing
precision makes it possible to test some fundamental relations in
cosmology, one of which is the well known Etherington's distance
duality relation (DDR) \citep{etherington1933on}. In the framework
of DDR, two cosmological distances, i.e., the luminosity distance
$D_L$ and angular diameter distance $D_A$ at the same redshift $z$
are connected as
\begin{equation}
\frac{D_L(z)}{D_A(z)}\lk(1+z\rk)^{-2} = 1.
\end{equation}
Theoretically, distances based on standard candles (e.g. supernovae)
and standard rulers (e.g. baryon oscillations) agree as long as
three conditions are met: the spacetime is characterized by a metric
theory, the light propagates along null geodesics, and the number of
photons is conserved. Despite of its application in all analyses of
cosmological observations, it is necessary to test the valid of this
relation because of the possible violation of the DD relation. It
has been argued in the literature that the violation of the former
two conditions is related to a signal of exotic physics acting as
the background gravity theory \citep{bassett2004cosmic}, while
cosmic opacity might contribute to the possibility that the number
of photons is not conserved in propagation
\citep{li2013cosmic,liao2015universe}. Therefore, a validity check
of the DDR not only tests the existing theories of gravity, but also
helps us understand some fundamental properties of the Universe.

Up to now, there are many works devoted to validate the DDR with
various observational data, in which the conclusions were drawn from
recent type-Ia supernovae data as luminous sources of known (or
standardizable) intrinsic luminosity in the Universe, while the
angular diameter distance was derived from different astrophysical
probes, such as Baryon Acoustic Oscillations (BAO)
\citep{wu2015cosmic,lv2016constraints,cardone2012testing,percival2010baryon,blake2011wigglez,beutler20116df},
Sunyaev-Zeldovich effect together with X-ray emission of galaxy
clusters
\citep{cao2011testing,bonamente2006determination,holanda2012test},
and strong gravitational lensing systems
\citep{biesiada2011dark,cao2012SL,cao2015SL,li2016comparison}, etc.
The first two tests are always con sidered as individual standard
rulers while the other two probes are treated as statistical
standard rulers in cosmology. However, it should be noted that the
information of angular diameter distance obtained from the above
standard rulers, is strongly model dependent, which will generate
systematic uncertainties hard to quantify \citep{liao2016distance}.
For instance, the angular diameter distance determined from X-ray
and SZ-studies of clusters is sensitive to the underlying geometry
of the galaxy clusters (spherical or elliptical)
\citep{holanda2012test} and the assumptions of the hot gas density
profile (simple $\beta$ or double $-\beta$ profile)
\citep{cao2016testing}. In addition, it was found that other
attempts using angular diameter distance data from Baryon Acoustic
Oscillations suffer from the limited sample size covering the
redshift range $0.35\leq z \leq 0.74$ \citep{yu2016new}. More
importantly, one note that with SN Ia and other standard rulers we
are only able to probe the relatively lower redshift range $z\leq
1.40$, which still remains challenging with respect to the
exploration of the DDR. In order to draw firm and robust conclusions
about the validity of DDR, one will need to minimize statistical
uncertainties by increasing the depth and quality of observational
data sets. In this paper, we will make a cosmological
model-independent test for the DD relation with two new methods by
using the simulated data of gravitational waves from the
third-generation gravitational wave detector (which can be
considered as standard siren to provide the information of
luminosity distance) and the newly-compiled sample of compact radio
quasars observed by very-long-baseline interferometry (which
represents a type of new cosmological standard ruler).

It is well known that the first direct detection of gravitational
waves (GWs) by the LIGO/Virgo collaboration
\citep{abbott2016observation} has opened the era of GW astronomy.
The original idea of using the waveform signal to directly measure
the luminosity distance $D_L$ to the GW sources (inspiralling and
merging double compact binaries) can be traced back to the paper of
\citet{schutz1986determining}, which indicates that the GW sources,
especially the inspiraling and merging compact binaries consisting
of neutron stars (NSs) and black holes (BHs), can also be used to
probe the information of the absolute luminosity distances. This
constraint from the so-called standard sirens originates from the
dependence of the $D_L$ measurements on the so-called chirp mass and
the luminosity distance \citep{schutz1986determining}. The
gravitational waves (GWs) provide an alternative tool to testing the
cosmology and there have been a number of attempts to do so
\citep{holz2005using,dalal2006short,zhao2011determination,taylor2012cosmology,cai2017estimating}.
The constraints derived in these works are compatible with
cosmological parameter constraints determined by other techniques
\citep{cao2011constraints,cao2012SL,cao2015SL}, if hundreds of GW
events are seen. More importantly, as a promising complementary tool
to supernovae, the greatest advantage of GWs lies in the fact that
the distance calibration of such standard siren is independent of
any other distance ladders. In this paper, we present a significant
extension of previous works and investigate the possibility of using
GW data to test the validity of DDR. Unfortunately, due to the
limited size of observed GW events up to date (which makes it
impossible to do statistical analysis), we will focus on a large
number of simulated GW events based on the third-generation GW
ground-based detector, Einstein Telescope (ET) \footnote{The
Einstein Telescope Project, https://www.et-gw.eu/et/}.

In the EM window, with the aim of acquiring angular diameter
distances, the angular size of the compact structure in radio
quasars provides an effective source of standard rulers in the
Universe. However, due to their uncertain intrinsic linear sizes, it
is controversial whether the compact radio sources can be calibrated
as standard cosmological probes
\citep{jackson1997deceleration,vishwakarma2001consequences,
lima2002dark,zhu2002cardassian,chen2003cosmological}. In general,
the precise value of the linear size $l_m$ might depend both on
redshifts and the intrinsic properties of the source, i.e., the
intrinsic luminosity $L$ \citep{gurvits1999measuring}. On the other hand, the
morphology and kinematics of compact structure in radio quasars
could be strongly dependent on the nature of the central engine,
including the mass of central black hole and the accretion rate
\citep{gurvits1999measuring}. Therefore, the central region may be standard if
these parameters are confined within restricted ranges for specific
quasars. Considering the possible correlation between black hole
mass $M_{BH}$ and radio luminosity $L_R$
\citep{laor2000on,jarvis2002on}, \citet{cao2016measuring} proposed
that intermediate-luminosity quasars ($10^{27}$ W/Hz $<L<10^{28}$
W/Hz) might be used as a new type of cosmological standard ruler
with fixed comoving length, based on a 2.29 GHz VLBI all-sky survey
of 613 milliarcsecond ultra-compact radio sources
\cite{kellermann1993cosmological,gurvits1994apparent}. More
recently, the value of $l_m$ was calibrated at $\sim 11$ pc though a
cosmological model-independent method \citep{cao2017ultra}, based on
which the cosmological application of the intermediate-luminosity
quasar sample (in the redshift range $0.46<z<2.76$) was also
extensively discussed in the framework of different dark energy
models \citep{li2017testing,zheng2017ultra} and modified gravity
theories \citep{qi2017new,xu2017new}. Compared with our previous
works focusing on SNe Ia as background sources
\citep{cao2011testing,holanda2012test}, the advantage of this work
is that, we achieve a reasonable and compelling test of DDR at much
higher redshifts ($z\sim 3.0$), which will help us to verify the
fundamental relations in the early Universe.

This paper is organized as follows. We briefly introduce how to
handle the gravitational wave data and the quasar data in Section
\ref{obs}. Then we show the analysis methods and results of our work
in Section \ref{result}. Finally, the conclusions and discussions
are presented in Section \ref{conclusion}.

\begin{figure}
\centering
\includegraphics[width=9cm,height=7cm]{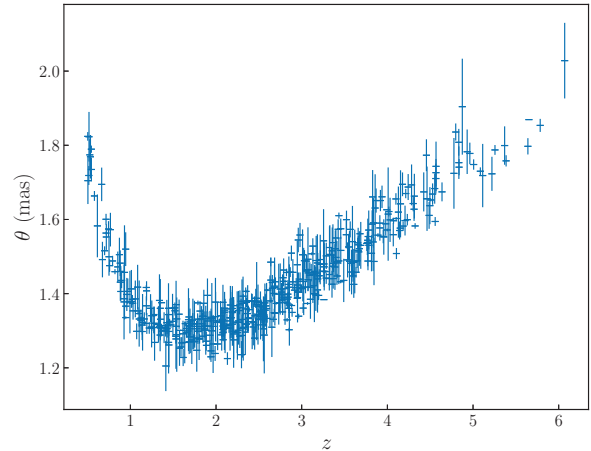}
\caption{The angular size measurements from 500 simulated radio
quasars.}\label{QSO_theta}
\end{figure}

\begin{figure}
\centering
\includegraphics[width=9cm,height=7cm]{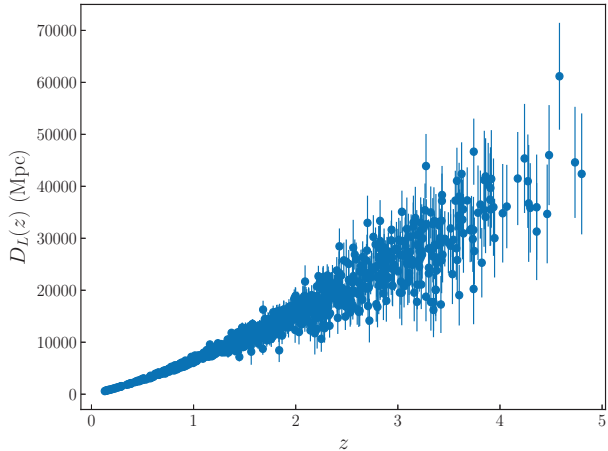}
\caption{The luminosity distance measurements from 1000 observed GW
events. }\label{DL}
\end{figure}

\section{Observations and simulations} \label{obs}

\subsection{Angular diameter distance from radio quasars}\label{qso}

Following the suggestions by Refs. \citep{gurvits1994apparent,vishwakarma2001consequences}, the linear
size of the compact structure in radio quasars depends on the
redshift $z$ and the intrinsic luminosity $L$ of the source
\begin{equation}
l_m=lL^{\beta}(1+z)^n
\end{equation}
where $\beta$ and $n$ are two parameters represent the "angular
size-redshift" and "angular size-luminosity" relations,
respectively. The parameter $l$ denotes the linear size scaling
factor describing the apparent distribution of radio brightness
within the core. In one of the more significant studies involving
the compact structure of radio quasars, \citet{gurvits1994apparent}
showed that the dispersion in linear size is greatly mitigated by
retaining only those sources with flat spectrum
($-0.38<\alpha<0.18$). These works have formed the basis of many
subsequent investigations
\citep{jackson1997deceleration,vishwakarma2001consequences,
lima2002dark,zhu2002cardassian,chen2003cosmological}. More recently,
based on a sample of 2.29 GHz VLBI survey with 613 milliarcsecond
ultra-compact radio sources covering the redshift range
$0.0035<z<3.787$, \citet{cao2016measuring} demonstrated that
intermediate-luminosity quasars (ILQSO), which show negligible
redshift and luminosity dependence $\lk(|n|\simeq
10^{-3},~\beta\simeq 10^{-4} \rk)$, may be used to establish some
general constraints on cosmological parameters. In the subsequent
analysis, the intrinsic linear size of this cosmological standard
ruler at 2.29 GHz was determined at \textbf{$l_m=11.03\pm0.25
\rm{pc}$,} which was obtained through a cosmological
model-independent method \citep{cao2017ultra}. This sub-sample
contains 120 intermediate-luminosity quasars covering the redshift
range $0.46<z<2.80$, whose cosmological application in $\Lambda$CDM
resulted with stringent constraints on both the matter density
$\Omega_m$ and the Hubble constant $H_0$, in a good agreement with
recent Planck 2015 results. It is now understood that these works
lead to a second significant advancement with the use of these
sources in the study of cosmic acceleration
\citep{li2017testing,zheng2017ultra,qi2017new,xu2017new}.

Let us remind that the angular sizes $\theta$ of these standard
rulers were estimated from the ratio of the correlated and total
flux densities measured with radio interferometers
($\Gamma=S_c/S_t$), i.e., the visibility modulus $\Gamma$ defines a
characteristic angular size
\begin{equation}
\theta={2\sqrt{-\ln\Gamma \ln 2} \over \pi B} \label{thetaG}
\end{equation}
where $B$ is the interferometer baseline measured in wavelengths
\citep{Thompson86}. It is reasonable to ask if the derived angular
sizes are dependent on the intrinsic luminosities of radio quasars.
At first sight this is the case, since they were obtained by
combining the measurements of total flux density $S_t$ and the
correlated flux density $S_c$ (fringe amplitude). However, one
should observe that in Eq.~(3) that the flux densities enters into
the angular size $\theta$ not through an $S$ measure directly, but
rather through a ratio of correlated and total flux densities
$\Gamma=S_c/S_t$. Therefore, the intrinsic luminosities of radio
quasars do not influence the derived angular sizes $\theta$, which
implies that the so-called "circularity problem" will not affect our
statistical analysis of the distance duality relation. Meanwhile, we
also remark here that the "$\theta$" value represents a
single-parameter Gaussian, which can be assumed to be a rough
representation of the complicated source structure (the actual
brightness distribution). The previous works convinced us
\citep{jackson2004tight,jackson2012ultra-compact} that we could
reliably define the size of an unresolved source though such
technique, while $\theta$ is accurate enough to represent source
sizes when averaged over a group of sources in a statistical
application. More importantly, besides the direct averaging over the
set of sources, it also mimics averaging over the different position
angles of the interferometer baselines, although the apparent
angular size of milliarcsecond structure in radio quasars is less
dependent on the orientation relative to the line of sight.
Following Ref. \citep{kellermann1993cosmological}, another useful
method to define the characteristic angular size of each source is
to measure the size between the peak in the map (i.e. the core) and
the most distant component exceeding a given relative brightness
level (i.e., 2\% of the peak brightness of the core), which was
extensively used in the investigation of compact structure in radio
sources \citep{Gurvits99,cao2015exploring}.

The observable of ILQSO is the angular size of the compact structure
$\theta$, which may then be written as
\begin{equation}
\theta(z)=\frac{l_m}{D_A(z)}.
\end{equation}
where $D_A(z)$ is the model-dependent angular diameter distance at
redshift $z$. Furthermore, we will consider the future observation
of radio quasars from VLBI surveys based on better uv-coverage which
will significantly reduce the uncertainty of the angular size of
compact structure observed. Consequently, one can expect to have a
better angular diameter distance information in the future, which
will allow us to test DDR more accurately. Here, in the simulation
below, we adopt the flat $\Lambda$CDM with $H_0=67.8
 \rm{km~s^{-1}Mpc^{-1}}$ and $\Omega_m=0.308$ based on the recent
Planck results \citep{ade2016planck}. Taking the linear size of
ILQSO as $l_m=11.03$ pc and following the redshift distribution of
QSOs \citep{palanque2016extended}, we have simulated 500 $\theta-z$
data in the redshift $0.50<z<6.00$, for which the error of the
angular size $\theta$ was taken at a level of $3\%$. This reasonable
assumption of the "$\theta$" measurements will be realized from both
current and future VLBI surveys based on better uv-coverage
\citep{pushkarev2015milky}. There are two general reasons for
ignoring sources with $z<0.5$. Firstly, as $z$ falls below 0.5, the
epoch of quasar formation comes to an end and the nature of the
population changes dramatically, which indicates the possible
existence of a correlation between linear size and radio luminosity.
Therefore, following the suggestion of Refs.
\citep{gurvits1994apparent,Gurvits99}, only the high-redshift part
of radio quasars could be used as a standard rod to fit different
cosmological models with experimental data. Secondly, as $z$
increases a larger Doppler boosting factor $\mathcal{D}$ is
required, i.e., the ratio $\mathcal{D}/(1+z)$ is approximately
fixed, so that the rest-frame emitted frequency
$(1+z)\nu_r/\mathcal{D}$ is also fixed. See Ref.
\citep{jackson2004tight} for mathematical and astrophysical details
\footnote{We remark here that, such approximation could possibly
constitute a source of systematic errors, i.e., it applies near the
redshift at which the angular diameter distance $D_A$ reaches its
maximum ($z\sim 1.5$ in the framework of $\Lambda$CDM cosmology),
and thus not at the high redshift regime investigated in this
work.}. Moreover, in order to make the simulated data more
representative of the experimental expectation, we assume the
angular size measurements obey the Gaussian distribution
$\theta_{mean}=\mathcal{N}(\theta_{fid},\sigma_{\theta})$ as shown
in Fig.~\ref{QSO_theta}. The more details of the specific procedure
of QSO simulation can be found in Ref. \citep{qi2018revised}.

\subsection{Luminosity distance from gravitation wave sources}\label{gw}

In this section we simulate GW events based on the Einstein
Telescope,the third generation of the ground-based GW detector.
Compared with the current advanced ground-based detectors (i.e., the
advanced LIGO and Virgo detectors), the ET is designed to be ten
times more sensitive covering the frequency range of $1-10^4$ Hz.
Here we briefly introduce the GW as standard sirens in the ET
observations.

GW detectors based on the ET could measure the strain $h(t)$, which
quantifies the change of difference of two optical paths caused by
the passing of GWs. It can be expressed as the linear combination of
the two polarization states
\begin{equation}
h(t)=F_+(\theta, \phi, \psi)h_+(t)+F_\times(\theta, \phi,
\psi)h_\times(t),
\end{equation}
where $F_{+,\times}$ are the beam pattern functions, $\psi$ denotes
the polarization angle, and ($\theta, \phi$) are angles describing
the location of the source relative to the detector. Following the
analysis of \citet{zhao2011determination}, the explicit expressions
of the beam patterns of the ET are given by
\begin{align}
F_+^{(1)}(\theta, \phi, \psi)=&~~\frac{{\sqrt 3 }}{2}[\frac{1}{2}(1 + {\cos ^2}(\theta ))\cos (2\phi )\cos (2\psi ) \nonumber\\
                              &~~- \cos (\theta )\sin (2\phi )\sin (2\psi )],\nonumber\\
F_\times^{(1)}(\theta, \phi, \psi)=&~~\frac{{\sqrt 3 }}{2}[\frac{1}{2}(1 + {\cos ^2}(\theta ))\cos (2\phi )\sin (2\psi ) \nonumber\\
                              &~~+ \cos (\theta )\sin (2\phi )\cos (2\psi )].
\label{equa:F}
\end{align}
Considering the fact that the three interferometers of the ET are
arranged in an equilateral triangle, the other two interferometer's
antenna pattern functions can also be derived from
Eq.~(\ref{equa:F}) \citep{cai2017estimating}
\begin{equation}
F_{+,\times}^{(2)}(\theta, \phi, \psi)=F_{+,\times}^{(1)}(\theta,
\phi+2\pi/3, \psi)
\end{equation}
\begin{equation}
F_{+,\times}^{(3)}(\theta, \phi, \psi)=F_{+,\times}^{(1)}(\theta,
\phi+4\pi/3, \psi).
\end{equation}

In this paper, we focus on the GW signals from the merger of binary
systems with component masses $m_1$ and $m_2$. Then the chirp mass
can defined as $\mathcal{M}_c=M \eta^{3/5}$, while the observed
counterpart can be written as $\mathcal{M}_{c,\rm
obs}=(1+z)\mathcal{M}_{c,\rm phys}$, where $M=m_1+m_2$ is the total
mass and $\eta=m_1 m_2/M^2$ represents the symmetric mass ratio.
Following
Refs.~\citep{sathyaprakash2009physics,zhao2011determination}, the
Fourier transform $\mathcal{H}(f)$ of the time domain waveform
$h(t)$ could be derived by applying the stationary phase
approximation,
\begin{align}
\mathcal{H}(f)=\mathcal{A}f^{-7/6}\exp[i(2\pi
ft_0-\pi/4+2\psi(f/2)-\varphi_{(2.0)})], \label{equa:hf}
\end{align}
where $t_0$ is the epoch of the merger, and the definitions of the
functions $\psi$ and $\varphi_{(2.0)}$ can be found in
\citep{zhao2011determination}. The Fourier amplitude $\mathcal{A}$
is given by
\begin{align}
\mathcal{A}=&~~\frac{1}{D_L}\sqrt{F_+^2(1+\cos^2(\iota))^2+4F_\times^2\cos^2(\iota)}\nonumber\\
            &~~\times \sqrt{5\pi/96}\pi^{-7/6}\mathcal{M}_c^{5/6},
\label{equa:A}
\end{align}
where $\iota$ represents the angle of inclination of the binary's
orbital angular momentum with the line of sight, and $D_L$ is the
theoretical luminosity distance in the fiducial cosmological model
we choose. It should be noted that the GW sources used in this work
are caused by binary merger of a neutron star with either a neutron
star or black hole, which can generate an intense burst of
$\gamma$-rays (SGRB) with measurable source redshift. More
specifically, since the SGRB is emitted in a narrow cone, a
criterion on the total beaming angle (e.g., $\iota<20^\circ$) should
be applied to detect one specific gravitational wave event
\citep{nakar2007short}. Moreover, as was pointed out in
\citet{li2015extracting,cai2017estimating}, averaging the Fisher
matrix over the inclination $\iota$ and the polarization $\psi$ with
the constraint $\iota<20^\circ$ is approximately equivalent to
taking $\iota=0$. Therefore, we can take the simplified case of
$\iota=0$ and then the Fourier amplitude $\mathcal{A}$ will be
independent of the polarization angle $\psi$. Given the waveform of
GWs, the combined signal-to-noise ratio (SNR) for the network of
three independent ET interferometers is
\begin{equation}
\rho=\sqrt{\sum\limits_{i=1}^{3}\left\langle
\mathcal{H}^{(i)},\mathcal{H}^{(i)}\right\rangle}. \label{euqa:rho}
\end{equation}
Here the inner product is defined as
\begin{equation}
\left\langle{a,b}\right\rangle=4\int_{f_{\rm lower}}^{f_{\rm
upper}}\frac{\tilde a(f)\tilde b^\ast(f)+\tilde a^\ast(f)\tilde
b(f)}{2}\frac{df}{S_h(f)}, \label{euqa:product}
\end{equation}
where $\tilde a(f)$ and $\tilde b(f)$ are the Fourier transforms of
the functions $a(t)$ and $b(t)$. $S_h(f)$ is the one-side noise
power spectral density (PSD) characterizing the performance of a GW
detector \citep{zhao2011determination}. The lower cutoff frequency
$f_{\rm lower}$ is fixed at 1 Hz. The upper cutoff frequency,
$f_{\rm upper}$, is decided by the last stable orbit (LSO), $f_{\rm
upper}=2f_{\rm LSO}$, where $f_{\rm LSO}=1/(6^{3/2}2\pi M_{\rm
obs})$ is the orbit frequency at the LSO, and $M_{\rm
obs}=(1+z)M_{\rm phys}$ is the observed total mass. Here, we
simulate many catalogues of NS-NS and NS-BH systems, with the masses
of NS and BH sampled by uniform distribution in the intervals of
[1,2] $M_{\odot}$ and [3,10] $M_{\odot}$. Meanwhile, the signal is
identified as a GW event only if the ET interferometers have a
network SNR of $\rho>8.0$, the SNR threshold currently used by
LIGO/Virgo network \citep{cai2017estimating}.

Moreover, using the Fisher information matrix, the instrumental
uncertainty on the measurement of the luminosity distance can be
estimated as
\begin{align}
\sigma_{D_L}^{\rm inst}\simeq \sqrt{\left\langle\frac{\partial
\mathcal H}{\partial D_L},\frac{\partial \mathcal H}{\partial
D_L}\right\rangle^{-1}},
\end{align}
if the uncertainty of $D_L$ is independent with the uncertainties of
the other GW parameters. Concerning the uncertainty budget,
following the strategy described by \citet{cai2017estimating}, the
distance precision per GW is taken as
$\sigma_{D_L}^2=(\sigma_{D_L}^{\rm inst})^2+(\sigma_{D_L}^{\rm
lens})^2$. In the simplified case of $\iota\simeq0$, the estimate of
the uncertainty of $D_L$ expresses as $\sigma_{D_L}^{\rm inst}\simeq
\frac{2D_L}{\rho}$ \citep{li2015extracting}, while the lensing
uncertainty caused by the weak lensing is modeled as
$\sigma_{D_L}^{lens}/D_L=0.05z$
\citep{sathyaprakash2010cosmography,li2015extracting,cai2017estimating}.
Thus, the total uncertainty of $D_L$ is taken to be
\begin{align}
\sigma_{D_L}&~~=\sqrt{(\sigma_{D_L}^{\rm inst})^2+(\sigma_{D_L}^{\rm lens})^2} \nonumber\\
            &~~=\sqrt{\left(\frac{2D_L}{\rho}\right)^2+(0.05z D_L)^2}.
\label{sigmadl}
\end{align}

Finally, we adopt the redshift distribution of the GW sources
observed on Earth, which can be written as
\citep{sathyaprakash2010cosmography}
\begin{equation}
P(z)\propto \frac{4\pi D_c^2(z)R(z)}{H(z)(1+z)}, \label{equa:pz}
\end{equation}
where $H(z)$ is the Hubble parameter of the fiducial cosmological
model, $D_c=\int_0^z1/H(z)dz$ is the corresponding comoving distance
at redshift $z$, and $R(z)$ represents the time evolution of the
burst rate taken as \citep{schneider2001low,cutler2009ultrahigh}
\begin{equation}
R(z)=\begin{cases}
1+2z, & z\leq 1 \\
\frac{3}{4}(5-z), & 1<z<5 \\
0, & z\geq 5.
\end{cases}
\label{equa:rz}
\end{equation}
The final key question required to be answered is: \emph{how many GW
events can be detected per year for the ET?} Focusing on the GW
sources caused by binary merger of neutron stars with either neutron
stars or black holes, recent analysis \citep{cai2017estimating}
revealed that the third generation ground-based GW detector can
detect up to 1000 GW events in a 10 year observation (with
detectable EM counterpart measurable source redshift). Therefore,
assuming the luminosity distance measurements obey the Gaussian
distribution $D_L^{mean}=\mathcal{N}(d_L^{\rm fid},\sigma_{d_L})$,
we simulate 1000 GW events used for statistical analysis in the next
section, the redshift distribution of which is shown in
Fig.~\ref{DL}.

\begin{table*}
\centering \setlength{\tabcolsep}{7mm}{
\begin{tabular}{l|ccc}
\hline \hline
Data & $\eta_0$ (Model I) &$\eta_1$ (Model II ) & $\eta_2$ (Model III) \\

QSO (Cur) + GW (Sim) [this work] & $-0.007\pm0.012$ &$-0.0086\pm0.0093$ & $-0.018\pm0.023$ \\
QSO (Sim) + GW (Sim) [this work] & $0.0002\pm0.0029$ &$-0.0004\pm0.0018$ & $-0.0007\pm0.0051$ \\
\hline
Union2 + galaxy cluster \citep{cao2011b} & $-0.03^{+0.05}_{-0.06}$ & $-0.01^{+0.15}_{-0.16}$ & $-0.01^{+0.21}_{-0.24}$ \\
Union2.1 + BAOs \citep{wu2015cosmic}& $-0.009\pm0.033$ & $0.027\pm0.064$  & $0.039\pm0.099$ \\
Union2.1 + $f_gas$ \citep{gonccalves2015constraints} & $\Box$ & $-0.08^{+0.11}_{-0.10}$ & $\Box$ \\
JLA + strong lensing \citep{liao2016distance}& $\Box$ & $-0.005^{+0.351}_{-0.215}$  & $\Box$ \\

\hline \hline
\end{tabular}}
\caption{Summary of the best-fit values for the DDR parameter
obtained from different observations. } \label{table1}
\end{table*}

\begin{figure}
\centering
\includegraphics[width=8cm,height=6cm]{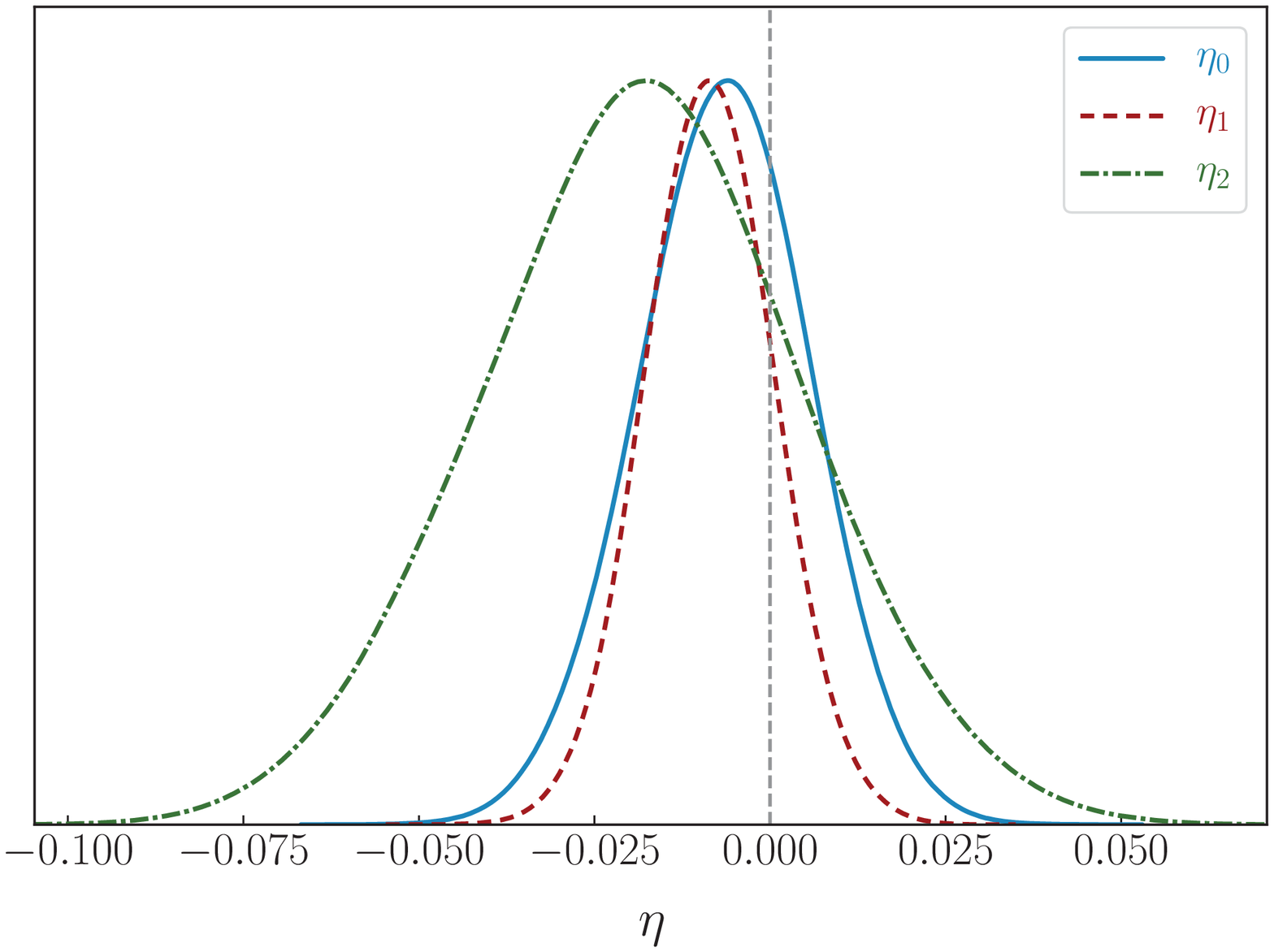}
\caption{The likelihood distributions of $\eta_0$, $\eta_1$ and
$\eta_2$ from the current QSO data and simulated GW data.
}\label{qso_gw}
\end{figure}

\begin{figure}
\centering
\includegraphics[width=8cm,height=6cm]{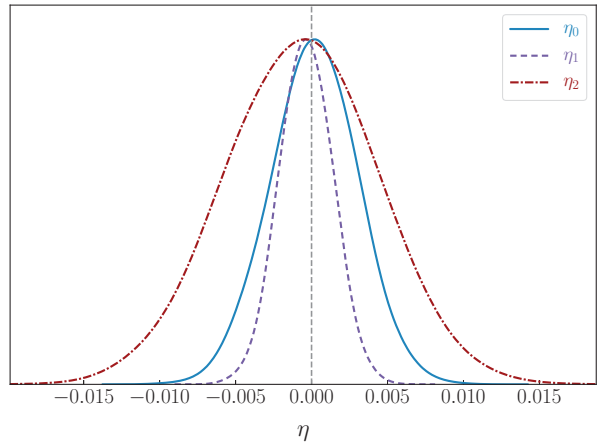}
\caption{The likelihood distributions of $\eta_0$, $\eta_1$ and
$\eta_2$ from the simulated QSO and GW data.}\label{future}
\end{figure}

\section{Constraints and results}\label{result}

From the theoretical point of view, in order to directly test the
DDR from GW+EM observations, our analysis will be based on one
constant parametrization and two parametric representations for
possible redshift dependence of the distance duality expression
\citep{li2011cosmological,holanda2012test}, namely
\begin{equation}
\eta(z)=\frac{D_L}{D_A}\lk(1+z\rk)^{-2},
\end{equation}
and
\begin{eqnarray}\label{equa:para}
\eta(z,\eta_0)&=&1+\eta_0, \\ \nonumber \eta(z,\eta_1)&=&1+\eta_1z,
\\ \nonumber \eta(z,\eta_2)&=&1+\eta_2\frac{z}{1+z}
\end{eqnarray}
where $\eta_0$, $\eta_1$ and $\eta_2$ are constant parameters, the
likelihood of which is expected to peak at zero in order to satisfy
the DD relation. Such parameterizations are clearly inspired on
similar expressions for the equation of state ($w$) in different
dark energy models \citep{cao2014cosmic}, i.e., the XCDM model
(where the equation of state parameter for dark energy is a
constant) and time-varying dark energy models (where the $w$
parameterizations stem from the first order Taylor expansions in
redshift $z$). Note that the first two expressions are continuous
and smooth linear expansion, while the last one may effectively
avoid the possible divergence at high redshifts. More importantly,
it should be noted that the deviations from DDR may point to a
non-metric spacetime structure (since the DDR depends only on
Lorentzian spacetime geometry apart from geometrical photon
conservation), which could possibly generate drastic effects on
light propagation \citep{Schuller:2017dfj}. The $\eta$
parameterizations in Eq. (\ref{equa:para}) are sufficient to account
for such non-metric behavior at leading orders, which is supported
by the recent discussion of the DDR on non-metric backgrounds
\citep{Schuller:2017dfj}.

From the observational point of view, for a given $D^{QSO}_A$ data
point, in order to check the validity of the DDR, the luminosity
distance from an associated GW data point $D^{GW}_L$ should be
observed at the same redshift. Following the recent analysis of
\citet{cao2011b}, the testing results of the DDR may be influenced
by the particular choice of the selection criteria for a given pair
of data set, i.e., the choice of $\delta z$ may play an important
role in this model-independent test. The redshifts of GW sample are
carefully chosen to coincide with the ones of the associated quasar
sample, which may hopefully ease the systematic errors brought by
redshift inconsistence between GW and EM observations. More
specifically, in our analysis, a selection criterion that bins $D_L$
measurements from GW within the redshift range
$|z_{QSO}-z_{GW}|<0.005$ is adopted to get $D_L$ at the redshift of
QSO. If $D_{\rm{L}i}$ represents the $i$th appropriate GW luminosity
distance with $\sigma_{D_{\rm{L}i}}$ denoting its reported
observational uncertainty, the weighted mean luminosity distance
$\bar{D}_L$ at the QSO redshift and its corresponding uncertainty
$\sigma_{\bar{D}_L}$ can obtained by the standard data reduction
framework \citep{bevington1993data}:
$\bar{D}_L=\frac{\sum\lk(D_{\rm{L}i}/\sigma^2_{D_{\rm{L}i}}\rk)}{\sum1/\sigma^2_{D_{\rm{L}i}}}$,
and $\sigma^2_{\bar{D}_L}=\frac{1}{\sum1/\sigma^2_{D_{\rm{L}i}}}$.
Subsequently, the observed $\eta_{\rm{obs}}(z)$ can be expressed as
\begin{equation}
\eta_{\rm{obs}}(z)=\frac{\bar{D}_L}{D_A}\lk(1+z\rk)^{-2},
\end{equation}
and the corresponding statistical error is given by
\begin{equation}
\sigma^2_{\eta_{\rm{obs}}}=\frac{\sigma^2_{\bar{D}_L}}{D_A^2}\lk(1+z\rk)^{-4}+\frac{\bar{D}_L^2}{D_A^4}\sigma^2_{D_A}\lk(1+z\rk)^{-4}.
\end{equation}
The likelihood estimator is determined by $\chi^2$ statistics
\begin{equation} \label{chi2}
\chi^2(\eta_j)=
\sum_i\frac{\lk[\eta(z,\eta_j)-\eta_{i,\rm{obs}}(z)\rk]^2}{\eta^2_{\eta_{i,\rm{obs}}}}
\end{equation}
for the three different parameterizations $j=0,1,2$.

To get reasonable $D_A$, we firstly turn to the recent catalog by
\citet{cao2017ultra} that contains 120 intermediate-luminosity
quasars, with redshifts ranging from 0.46 to 2.80, all observed with
Very Large Baseline Interferometry (VLBI). Considering the
uncertainties in $D_A$ encountered previously
\citep{cao2016measuring}, we include the statistical error of
observations in $\theta(z)$ and an additional 10\% systematical
uncertainty accounting for the intrinsic spread in the linear size.
The GW data are carefully selected whose redshift is closest to the
quasar's redshift, demanding that the difference in redshift is
smaller than 0.005. Combining these quasar data together with the GW
estimate of the luminosity distances, for Model I we obtain the
best-fit value $\eta_0=-0.007\pm0.012$ at 68.3\% confidence level
and plot the likelihood distribution function in Fig.~3. Working on
the other two parameterization forms of the DD relation:
$\eta(z)=1+\eta_1z$ and \textbf{$\eta(z)=1+\eta_2z/(1+z)$}, the
best-fit values are $\eta_1=-0.0086\pm0.0093$ and
$\eta_2=-0.018\pm0.023$ at 68.3\% confidence level. The results are
summarized on Table I and are depicted on Fig.~3, which indicate
that $\eta<1$ tends to be slightly favored by all three
parametrizations of $\eta(z)$. Such tendency has been previously
noted and extensively discussed in the literature
\citep{cao2011b,gonccalves2015constraints,liao2016distance}. We
remark here that, compared with the previous works based on
observations of $D_A$ on large angular scales (galaxy clusters
\citep{cao2011b,li2011cosmological,gonccalves2015constraints}, BAO
\citep{wu2015cosmic}, galaxy strong lensing systems
\citep{liao2016distance}), using such different technique (compact
structure in radio quasars) to estimate $D_A$ opens the interesting
possibility to test the fundamental relations in the early universe
($z\sim 3$). Moreover, it is necessary to compare our results with
those of earlier studies using alternative probes at high redshifts
(GRBs and SGL systems). More recently, \citet{Yang2017} tested the
DD relation with current strong lensing observations
\citep{cao2015SL} and future luminosity distances from gravitational
waves sources, with the final conclusion that the DD relation can be
accommodated at 1$\sigma$ (C.L.). In our analysis, in the framework
of model-independent methods testing the DDR, the current compiled
quasar sample may achieves constraints with much higher precision of
$\Delta \eta=10^{-2}$.


On the other hand, we also pin our hope on the VLBI observations of
more compact radio quasars with higher angular resolution based on
better uv-coverage. In order to compare with previous results from
the current quasar sample, we also derive the testing results from
simulated QSO and GW data in Table I, with the best-fit values of
the $\eta$ parameter in the three DDR models:
$\eta_0=0.0002\pm0.0029$ for Model I, $\eta_1=-0.0004\pm0.0018$ for
Model II, and $\eta_2=-0.0007\pm0.0051$ for Model III. The
corresponding likelihood distribution function from three
one-parameter forms of DDR parameterizations are also shown in
Fig.~\ref{future}. Furthermore, the future VLBI observations of
ILQSO combined with the simulated data of GWs using the Einstein
Telescope (ET) could extend the test of DDR to much higher redshifts
(i.e., $z\sim 5$). More importantly, one can clearly see that the
future compiled quasar data improves the constraints on model
parameters significantly. With the confrontation between the angular
diameter distance (ADD) from quasars and luminosity distance (LD)
from GWs, one can expect the validity of the distance duality
relation to be confirmed at the precision of $\Delta \eta=10^{-3}$.

Now it is worthwhile to make some comments on the results obtained
above. As was commented earlier, the cosmic opacity caused by the
absorption or scattering effects of dust in the Universe might
contribute to the possible violation of DDR. In particular, the
latest observations of SN Ia, which strongly support the accelerated
expansion of the Universe may be affected by the dust in their host
galaxies and Milky Way
\citep{li2013cosmic,liao2015universe,qi2018what}. However, it should
be emphasized that the luminosity distance derived from waveform and
amplitude of the gravitational waves observations is insensitive to
non-conservation of the number of photons \citep{Yang2017}.
Therefore, the method proposed in our analysis opens an interesting
possibility to probe exotic physics in the theory of gravity
\citep{bassett2004cosmic}, as can be seen from possible deviation
from the standard distance duality relation.

\section{Conclusions} \label{conclusion}

In this paper, we have discussed a new model-independent
cosmological test for the distance duality relation. For $D_L$ we
consider the simulated data of gravitational waves from the
third-generation gravitational wave detector (ET), which can be
considered as standard siren, while the angular diameter distances
$D_A$ are derived from the newly-compiled sample of compact radio
quasars observed by VLBI, which represents a type of new
cosmological standard ruler. This creates a valuable opportunity to
directly test DDR at much higher precision with the combination of
gravitational wave (GW) and electromagnetic (EM) signals. In order
to obtain a more reliable result of testing the DDR from GW+EM
observations, we use one constant parametrization
($\eta(z)=1+\eta_0$) and two parametric representations for possible
redshift dependence of the distance duality expression
($\eta(z)=1+\eta_1z$, \textbf{$\eta(z)=1+\eta_2z/(1+z)$)}. The
redshifts of GW sample are carefully chosen to coincide with the
ones of the associated quasar sample, which may hopefully ease the
systematic errors brought by redshift inconsistence between GW and
EM observations. More specifically, in our analysis, a selection
criterion that bins $D_L$ measurements from GW within the redshift
range $|z_{QSO}-z_{GW}|<0.005$ is adopted to get $D_L$ at the
redshift of QSO.

Firstly of all, we turn to the recent catalog by
\citet{cao2017ultra} that contains 120 intermediate-luminosity
quasars with redshifts ranging from 0.46 to 2.80, all observed with
Very Large Baseline Interferometry (VLBI). Combining these quasar
data together with the GW estimate of the luminosity distances, we
obtain the best-fit value $\eta_0=-0.007\pm0.012$,
$\eta_1=-0.0086\pm0.0093$ and $\eta_2=-0.018\pm0.023$ at 68.3\%
confidence level, which indicate that $\eta<1$ tends to be slightly
favored by all three parametrizations of $\eta(z)$. In the framework
of model-independent methods testing the DDR, the current compiled
quasar sample may achieve constraints with much higher precision of
$\Delta \eta=10^{-2}$. Moreover, compared with the previous works
based on observations of $D_A$ on large angular scales (galaxy
clusters, BAO, galaxy strong lensing systems), using such different
technique (compact structure in radio quasars) to estimate $D_A$
opens the interesting possibility to test the fundamental relations
in the early universe ($z\sim 3$). Therefore, the spacetime
characterized by a metric theory and that the light propagates along
null geodesics is strongly supported by the available observations.
This is the most unambiguous result of the current data sets.

Working on more simulated compact radio quasars with higher angular
resolution based on better uv-coverage, our results show that the
future VLBI observations of ILQSO combined with the simulated data
of GWs using the Einstein Telescope (ET) could extend the test of
DDR to much higher redshifts (i.e., $z\sim 5$). More importantly,
one can expect the validity of the DDR to be confirmed at the
precision of $\Delta \eta=10^{-3}$. Since the luminosity distances
obtained from GW observations are insensitive to the
non-conservation of photon number, any deviation from the standard
distance duality relation can be explained as possible existence of
exotic physics in the gravity theory. This encourages us to expect
the possibility of testing DDR at much higher precision in the
future, which reinforces the interest in the observational search
for more quasar samples and GW events with smaller statistical and
systematic uncertainties.

\acknowledgments{ This work was supported by National Key R\&D
Program of China No. 2017YFA0402600; the Ministry of Science and
Technology National Basic Science Program (Project 973) under Grants
No. 2014CB845806; the National Natural Science Foundation of China
under Grants Nos. 11503001, 11690023, 11633001, and 11873001;
Beijing Talents Fund of Organization Department of Beijing Municipal
Committee of the CPC; the Fundamental Research Funds for the Central
Universities and Scientific Research Foundation of Beijing Normal
University; and the Opening Project of Key Laboratory of
Computational Astrophysics, National Astronomical Observatories,
Chinese Academy of Sciences. J.-Z. Qi was supported by the China
Postdoctoral Science Foundation (Grant No. 2017M620661). J. Li was
supported by the national natural science fund of china Grant No. 11873001.
Y. Pan was supported by CQ CSTC under grant Nos. cstc2015jcyjA00044 and
cstc2018jcyjAX0192, and CQ MEC under grant No. KJ1500414.}

\bibliography{ddr}

\end{document}